# Consortium Blockchain for Security and Privacy-Preserving in E-government Systems
*(Full Paper)*

Noe Elisa, Northumbria University, noe.nnko@northumbria.ac.uk
Longzhi Yang*, Northumbria University, longzhi.yang@northumbria.ac.uk
Honglei Li, Northumbria University, honglei.li@northumbria.ac.uk
Fei Chao, Department of Computer Science, Aberystwyth University, fec10@aber.ac.uk
Nitin Naik, Defence School of Communications of Information Systems, Ministry of Defense, UK, nitin.naik100@mod.gov.uk

**ABSTRACT**

Since its inception as a solution for secure cryptocurrencies sharing in 2008, the blockchain technology has now become one of the core technologies for secure data sharing and storage over trustless and decentralised peer-to-peer systems. E-government is amongst the systems that stores sensitive information about citizens, businesses and other affiliates, and therefore becomes the target of cyber attackers. The existing e-government systems are centralised and thus subject to single point of failure. This paper proposes a secure and decentralised e-government system based on the consortium blockchain technology, which is a semi-public and decentralised blockchain system consisting of a group of pre-selected entities or organisations in charge of consensus and decisions making for the benefit of the whole network of peers. In addition, a number of e-government nodes are pre-selected to perform the tasks of user and transaction validation before being added to the blockchain network. Accordingly, e-government users of the consortium blockchain network are given the rights to create, submit, access, and review transactions. Performance evaluation on single transaction time and transactions processed per second demonstrate the practicability of the proposed consortium blockchain-based e-government system for secure information sharing amongst all stakeholders.

*Keywords*: E-government systems, Consortium blockchain, Validators, Security, Privacy, Decentralised systems.

\_\_\_\_\_\_\_\_\_\_\_\_\_\_\_\_\_\_\_\_\_\_\_\_
*Corresponding author

**INTRODUCTION**

The rapid advancement of Information and Communication Technologies (ICT) has made it possible to store and access vast amount of information effectively and efficiently in networked systems. E-government is one such system that uses ICT to deliver public services to citizens, businesses and other stakeholders. The transformation of government systems from traditional paper-based information sharing to its electronic counterpart increases the transparency, accountability, participatory, effectiveness and efficiency of services delivered by government agencies, resulting into a better government system (UN, 2018). In the future, the number of devices using e-government systems will increase dramatically due to the proliferation of technologies such as smart cities, internet of things (IoTs), cloud computing and interconnected networks (Yang et al., 2019; Elisa et al., 2018b). Commonly, as the number of devices using e-government systems rises, the number of malicious nodes trying to access and abuse unauthorised and sensitive information grows accordingly (Yang et al., 2019). The goal of an e-government system is to ensure that information security goals including integrity, confidentiality, availability and trust are met for its users. Thus, for the successful adoption of e-government systems, cyber security provision is of paramount importance (Lambrinoudakis et al., 2003; Elisa, 2017).

Data breaches in e-government systems have been significantly increasing in recent years based on various sources. For instance, according to the 2019 Cyber Security Breaches Survey by the UK government, around 32% of businesses and 22% of charities reported facing cyber security breaches or attacks in 2019, such as phishing, viruses, malware including ransomware attacks and impersonation of emails (UK Cyber-Attacks, 2019). In the United States of America, more than 5.90 billion data have been stolen by cyber criminals so far in 2019 (US Cyber-Attacks, 2019), including social security numbers, credit card numbers, names, passwords and addresses. Recently, 364 million Chinese citizens' sensitive information were leaked online by unknown hackers, revealing private information like user name, photos, address and identity card numbers (China Attacks, 2019). Another recent cyber security breaches in e-government systems happened in Tanzania where the country was hit by hackers and cyber-terrorists who caused a loss of around 85 million US dollars (IPP Media, 2018).

There is a demand for efficient and robust mechanisms to ensure information security and data privacy, due to the centralised nature of the existing e-government systems. Blockchain technology provides the decentralised environment for secure transaction processing in trustless systems. It is designed as an immutable and distributed database for protecting privacy and security of the shared transactions among its trustless participants. In fact, the blockchain technology has been successfully exploited for security and privacy provision in supply chain (Tian, 2016), healthcare system (Peterson et al., 2016), Internet of Things (IoT) (Dorri et al., 2017), land registry (Ramya et al., 2018), smart cities (Biswas and Muthukkumarasamy, 2016), educational systems (Turkanovi´c et al., 2018), in addition to the well-known e-currency. The blockchain technology can be public (permissionless), private (permissioned) or consortium (semi-public and semi-private). One type of blockchain technology that is designed to meet the need of the enterprise is the consortium blockchain. The consortium blockchain technology is a semi-public and decentralised blockchain system which consists of a group of pre-selected entities or organisations responsible for consensus and decisions making for the benefit of the whole network of peers.

This study proposes a secure and decentralised e-government system based on the consortium blockchain technology. Briefly, the proposed consortium blockchain-based e-government system consists of four layers namely the services access layer, consortium blockchain layer, network layer and ledger storage layer. The service access layer provides services to e-government users by connecting their devices to the consortium blockchain-based e-government system. The consortium layer includes a number of pre-selected e-government nodes for validating users and transactions before being added to the blockchain network. The network layer ensures the connectivity between different entities of the proposed systems. The ledger storage layer stores off-chain (sideDB) data such as images, PDFs, DOCs, contracts, and other files that are too large to be stored in the blockchain or that are required to be deleted or changed in the future. The performance evaluation based on the number of transactions processed per second and on the time for processing a single transaction by varying the number of validators in the consortium blockchain network have proved that, the proposed system is suitable for security and privacy assurance in e-government systems.

The remainder of this paper is organised into four sections namely background, consortium blockchain-based e-government system, performance evaluation and finally conclusion and future direction as follows.

## BACKGROUND
This section details the background on the blockchain technology and e-government systems.

**Blockchain Technology**
The blockchain technology is a decentralised peer to peer (P2P) network that maintains a list of continuously growing shared database (ledger) running on the Internet. The transactions are chained together to form a "block" of records, thus named as blockchain. Each participant of the blockchain network has a pair of private and public key for signing and verifying transactions (Nakamoto, 2008). SHA-256 Cryptographic Hash Algorithm (Wolrich et al., 2014) is a commonly used to generate the hash value from the public key of each participant, which is then used as the participant's identity. The participants of the blockchain network use the distributed consensus algorithms such as Prof of Work (PoW) (Nakamoto, 2008), Proof of Stake (PoS) (Mingxiao et al., 2017), Delegated PoS (Mingxiao et al., 2017) or Byzantine Fault Tolerance (Mingxiao et al., 2017), to agree on the validity and the order in which transactions are added to the blockchain network.

A blockchain ledger in illustrated in Figure 1. The first block known as the genesis is formed by hardcoding and embedding random transactions into the blockchain software (Nakamoto, 2008). The upper part of each block contains its header and the lower part contains the list of transactions (body). The header includes the cryptographic hashes of the current and previous block headers linked together to create a secure chain of blocks. The header also stores the timestamp, a nonce, block version, and the root hash of all transactions stored in that block. The timestamp is used to indicate the time taken (representing the difficulty) to mine and add the block. The block version indicates the validation rules followed to create the block. The nonce is a 4-byte unique random number which normally starts with 0 and increases for every next transaction. The root hash of the transactions is derived from a Merkle tree as shown in the lower part of Figure 1. In a Merkle tree, each leaf node contains a hash of a single transaction, and all non-leaf nodes store the hashes of their children. A typical Merkle tree consists of two (2) branches, making each node hold up to two children. The Merkle root summarises all transactions in the block so as to maintain the integrity of data in case of tempering.

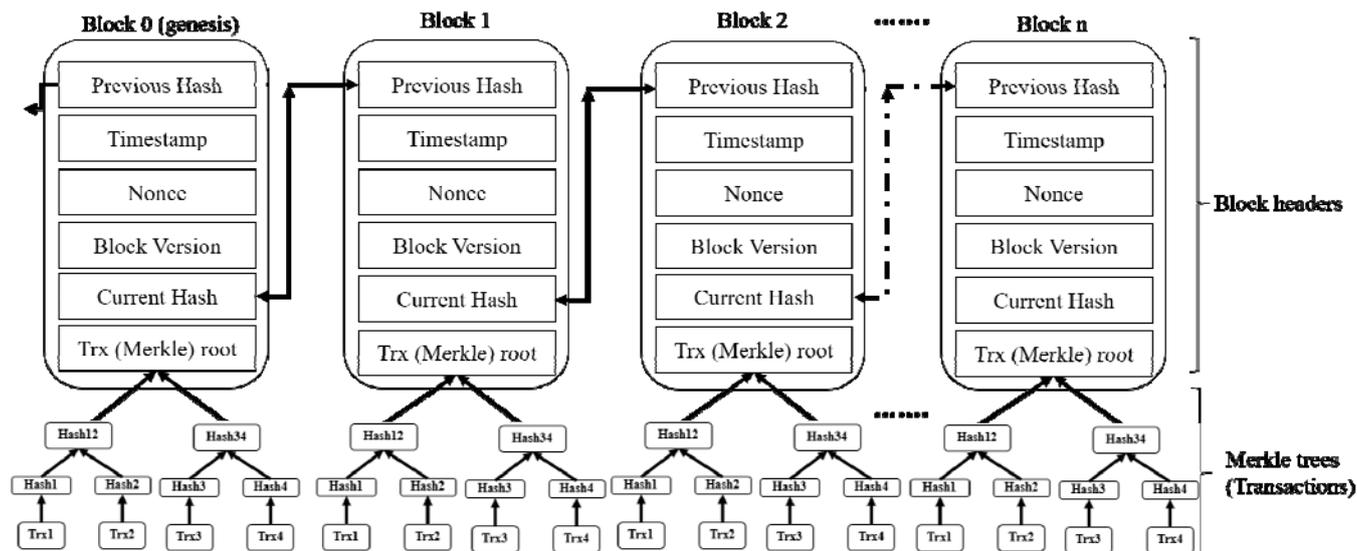
Figure1: An example of a detailed blockchain ledger.

Example applications of blockchain technologies include Bitcoin (support decentralised cryptocurrencies) (Nakamoto, 2008), Ethereum (support self-executing digital smart contracts) (Buterin et al., 2014), Hyperledger Fabric (support development of general enterprise solution) (Cachin, 2016), amongst others. Typically, the blockchain technology can be either public (permissionless), private (permissioned) or consortium (semi-public and semi-private) as detailed below.

*Public blockchain:* In a public blockchain, any individual can view, modify, and audit the blockchain without having a single entity in charge of the whole network. The consensus and decision making is reached through a decentralised consensus manner such as PoW in bitcoin (Nakamoto, 2008). The computation power of the participants of the blockchain network is used to select one participant powerful enough to add new transitions to the distributed ledger. The participants are incentivised every time when adding new transactions to the blockchain network, and thus this motivates everyone to use more computations to get the chance of adding transaction to the ledger. In public blockchain network, the higher the number of users, the more secure the network; as it creates a network of trusted individuals between the participants.

*Private blockchain:* The private blockchain is owned by an individual organisation who is responsible for granting access to the network for new users. Only few specific individuals in the origanisations have rights to validate transactions and blocks, and append them to the blockchain network. It is centralised compared to the public blockchain.

*Consortium blockchain:* The consortium blockchain consists of a pre-selected set of nodes or computers that are responsible for controlling access to the blockchain network resources (Dib et al., 2018). The goal of the consortium blockchain is to eliminate the individual/single autonomy of the private blockchain by having multiple entities or organisations in charge of consensus and decision making for the benefit of the whole network of peers. Since only pre-selected organisations are allowed to validate transactions and consensus, incentives are not necessary in this network. The pre-selected set of nodes make it partially private, partially public and semi-decentralised. More precisely, it provides the benefits of public blockchain in terms of efficiency and scalability while still permit some degree of central safeguarding and monitoring like in private blockchain. The consortium blockchain such as Hyperledger fabric (Cachin, 2016) is designed to meet the needs of the enterprises where a group of collaborating agencies exploit the blockchain technology to improve service delivery. All consensus participants of the consortium blockchain are known and reputable, therefore, malicious users cannot join the network freely.

**E-government Systems**
E-government systems use ICT to provide friendly services to citizens and businesses across different government departments. The use of e-government web portals enables organisations in private and public sectors to interact directly regardless of their physical locations around the world, whilst simultaneously improving the quality, convenience, transparency, efficiency and effectiveness of public services. It is becoming a mandatory for most countries across the world to use digital communication between citizens, businesses and the government (UN, 2018). Additionally, the use of e-government makes public administrators and officials more democratic and responsible due to the provided tractability and transparency.
E-government systems are categorised into four groups including Government to Citizens (G2C), Government to Business (G2B), Government to its Employees (G2E), and Government to Government (G2G) (Karokola, 2012). G2C is the interaction between citizens and government via government web portals. G2B involves communication between the government and business partners or other corporate organisations to share information such as procurements, company registration and payment for licenses and taxes. G2E can be referred to as intra-government communication concerned with the sharing of the documents among employees of the government. G2G is the interactions with other governments as well as the internal communication between government agencies or department by following the established rules governing public services delivery.

The existing technologies for delivering e-government services to citizens and other stakeholders include web portals, mobile phones (sms and mobile applications), and electronic identity (eID) (Stefanova et al., 2006). E-government web portal is an integrated application that serves as an entry to government websites to provide individuals direct access to online services provided by the government. Mobile government uses mobile phones within public administrations to deliver services to individuals. The eID is the digital identity which is assigned to citizen or an organisation to help the government to authenticate legitimate users (Stefanova et al., 2006). An individuals can use the same eID to access to different services offered by different public departments. eID solution can also be provided using mobile technologies (Stefanova et al., 2006). Common to these technologies is that, information collected from individuals are stored in centralised databases and servers. The centralised management systems can become a single point of failure or target to malicious users. Numerous artificial intelligence techniques such as (Yang et al., 2017; Elisa et al., 2018) have been exploited to develop intrusion detection systems (IDSs) that can also be used by e-government systems to enforce security, but they cannot be used to preserve privacy neither ensure decentralisation (Elisa et al., 2018b).

## CONSORTIUM BLOCKCHAIN-BASED E-GOVERNMENT SYSTEM

The proposed consortium blockchain-based e-government architecture is illustrated in Figure 2. It consists of four layers including services access layer, consortium blockchain layer, network layer and the ledger storage layer. The services access layer comprises of e-government users and various devices for providing access, computing resource, and storage of user's credentials. The consortium blockchain layer is a P2P network of pre-selected e-government validators responsible for validating transactions and authenticating e-government users before joining the network. Also, the consortium blockchain layer is responsible for provision of various services to users including consensus, P2P communication, and users' identity management. The network layer is for providing the connectivity between users, e-government consortium blockchain network and the ledger storage. The ledger storage layer is for storing off-chain (sideDB) data such as images, PDFs, DOCs, contracts, and other files that are too large to be stored in the blockchain or that are required to be deleted or changed in the future. The requirement of the ledger storage layer is because the blockchain database is append-only and immutable, so there is no chance of deletion in the future. Each of these layers is detailed in the following subsections.

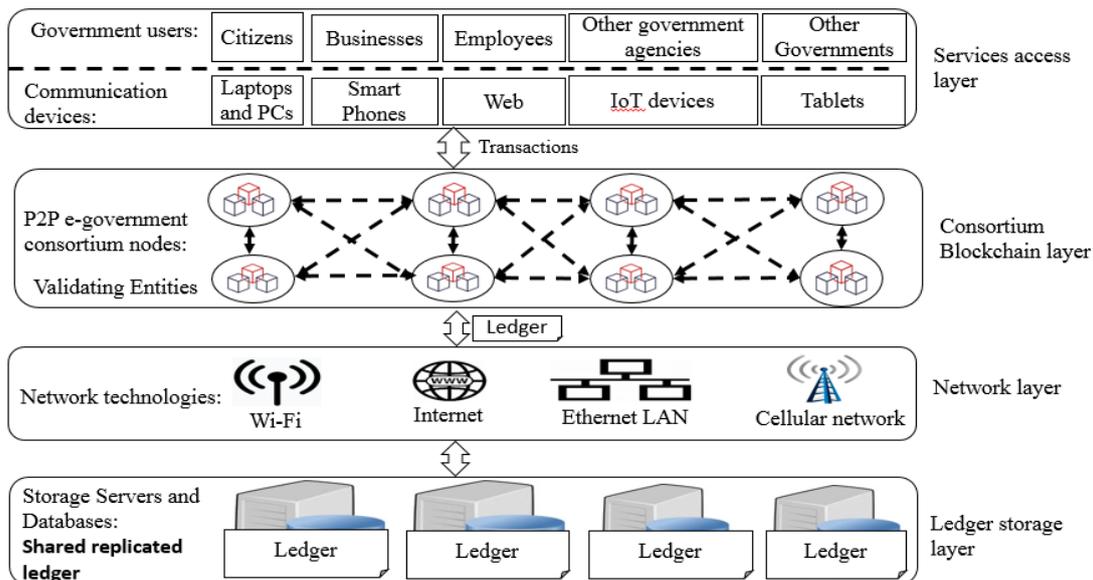

Figure 2: The proposed consortium blockchain e-government system.

**Service Access Layer**

This layer consists of e-government user information and various devices for providing access, computing resource, and storage of credential data. User's devices on this layer can be PCs, laptops, smart phones, tables, through which they can write or read data to the e-government consortium blockchain network. The devices at this layer will use blockchain interface embedded on web browser or mobile application to interact with the blockchain network of government consortium nodes.

Initially, when an e-government user signs up for government services, a private key, public key, identity and the corresponding blockchain address will be generated and stored in his/her wallet as illustrated in Algorithm 1. Note that, the blockchain wallet will be generated as soon as the user signs up with e-government system. The private key is for signing and authorising transactions where verification by other blockchain entities is done by using the corresponding public key.

Once an e-government user wants to access the service, the request is signed by the private key and get broadcast to the consortium blockchain of government peers. A copy of account information and transactions history for every user will be stored in the e-government consortium servers to facilitate account recovery in case the user loses the wallet password.

```
Algorithm 1 User Registration
input: New user request
output: Registered user u
 1: (K_pub, K_pr) ← keysGeneration();
 2: ID ← userIDGeneration();
 3: wallet ← generateWallet() + (K_pub, K_pr);
 4: BlockAddress ← generateAddress() + (K_pub, K_pr);
 5: (ID, K_pr, K_pub) ← walletStore(uID, K_pr, K_pub);
 6: storeWallet(wallet);
 7: u ← registeredUser();
```

**Consortium Blockchain Layer**

The participants of this layer are the pre-selected government departments (validators), who are responsible for validating transactions and authenticating e-government user's registration with the consortium blockchain network. The departments that meet the requirements for pre-selected participants of the consortium blockchain network will be registered by using Algorithm 2. After registration, a node will receive a blockchain address, private and public key pair for signing and verifying transactions.

```
Algorithm 2 Adding pre-selected peer to the consortium network
input: A new participant node,
       A set of N nodes in the current consortium network
output: Registered node m
 1: (K_pub, K_pr) ← keysGeneration()
 2: ID ← peerIDGeneration();
 3: wallet ← generateWallet() + (K_pub, K_pr);
 4: BlockAddress ← generateAddress() + (K_pub, K_pr);
 5: (ID, K_pr, K_pub) ← walletStore(uID, K_pr, K_pub);
 6: storeWallet(wallet);
 7: for each n ∈ {N − m} do
 8:    broadcastRegistration(n, m);
 9: end for
10: m ← verifiedValidator();
```

A number of validators amongst the consortium peers are required to validate any transaction before being added to the blockchain ledger. For instance, the consensus could be reached if a transaction is verified by twenty departments in a consortium blockchain with fifty departments in total. Thus, the validators collectively sign off transactions, grant or deny access to the shared blockchain ledger. Note that, the ledger in this layer is permissioned since only approved participants of the blockchain network have the access privilege. Any change to the blockchain stored in the consortium layer will be reflected by all participants in their copies as soon as it is realised. Additionally, the consortium blockchain layer provides e-government users the rights to create and submit transactions, access, and review their transactions.

The process for creating and adding a block to the blockchain by the validators is illustrated in Algorithm 3.

```
Algorithm 3 Adding a new block to the consortium blockchain
input: A set of N participants in the consortium blockchain network,
       A blockchain B including blocks from the genesis to b_n,
       Validation time T_c needed to create a block
output: A new block b_{n+1}
        A new blockchain B'
 1: Initialise an empty set of transaction R = {};
 2: ω ← electValidators(N);
 3: while validation_time < T_c do
 4:    for each n ∈ {N − ω} do
 5:       R ← R + addTransactionfromUser(n);
 6:    end for
 7: end while
 8: b_{n+1} ← newBlockCreation(b_n, R);
 9: for each n ∈ {N − ω} do
10:    signnewBlock(b_{n+1}, n);
11: end for
12: B' ← B + b_{n+1};
13: for each n ∈ N do
14:    broadcastBlockchain(B', n);
15: end for
```

The boundaries formed by the consortium layer of blockchain nodes and e-government users is illustrated in Figure 3. The communication between the consortium blockchain network and the services access layer happens on demand when users submit transactions.

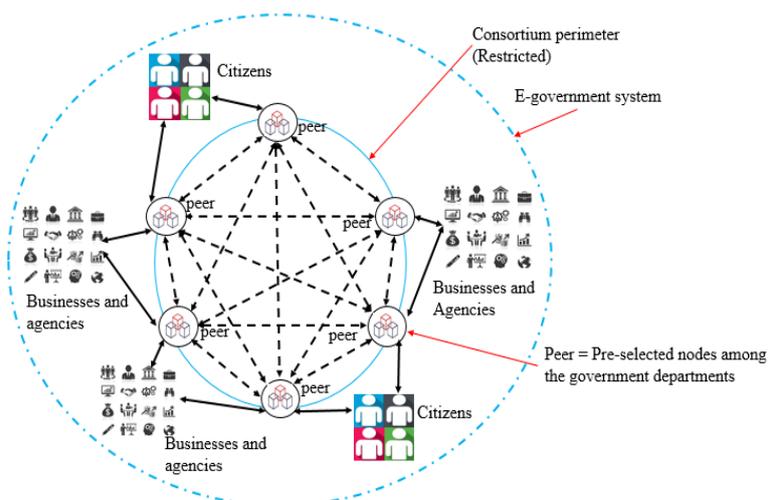

Figure 3: The consortium blockchain-based e-government network.

**Network Layer**
The main function provided by the network layer is the connectivity between e-government users, government consortium blockchain layer and the ledger storage servers. There is a wide choice of technologies that can be used on this layer such as WiFi, Ethernet LAN or cellular network. The participants of the consortium blockchain will use this layer to communicate transactions to/from the users and store new blocks in the ledger storage layer. Wireless communication provided by this layer will be used to support wireless devices whilst interacting with the e-government consortium blockchain network, because many governments around the globe are trying to incorporate wireless broadband networks across different cities using WiFi technology due to its great flexibility and convenience (Wireless Broadband, 2018).

**Ledger Storage Layer**
This layer is introduced in order to enable storage and replication of government and users data. This layer will be used to store off-chain (sideDB) data such as pictures, contracts, PDFs, and other files that are too large to be stored in the blockchain or that are required to be deleted or changed in the future. The blockchain database is append-only and immutable; so they cannot be deleted in the future. This layer is so critical in case that data reuse is required as the data that are store in the blockchain network are immutable and difficult to be reused. The off-chain storage of data is required for documentation and verification in blockchain technology (Cachin, 2016). A hash value for the off-chain document will be produced and stored in the consortium blockchain ledger whilst the real document is stored in the ledger storage layer. Additionally, the data that are processed by the participants of the consortium blockchain will be replicated and stored in this layer to allow more storage spaces for the blockchain network devices.

## PERFORMANCE EVALUATION
A detailed technical analysis on the performance of the consortium blockchain is reported in (Dib et al., 2018), which helps organisations to evaluate the system before its adoption. The same performance measures, including the single transaction time and the number of transactions validated per second, were also utilised in this work, to evaluate the proposed consortium blockchain-based e-government system. All technical evaluations were performed by using the Ethermint protocol of Ethereum blockchain which support smart contracts over consortium network (Dib et al., 2018).

**Performance Based On The Number Of Transactions Per Second**
The performance of the consortium blockchain network based on the number of transactions processed per second with increasing number of validators in the network, is illustrated in Figure 4. Note that, the ideal case happens when all transactions from the users are validated in one second or less. It is clear from the figure that, the number of average transactions validated per second decreases as the number of validators increases in the consortium blockchain network. This is because the communication overhead required to pre-vote for a node will process the transactions, create a block, and append it to the ledger. Thus, if a given consortium blockchain system requires a large number of validators, the transaction processing speed will decrease. In e-government systems, this may not be the case since all participating department save the same goal of proving public services, fewer validators is usually enough and trusted to process and validate transactions.

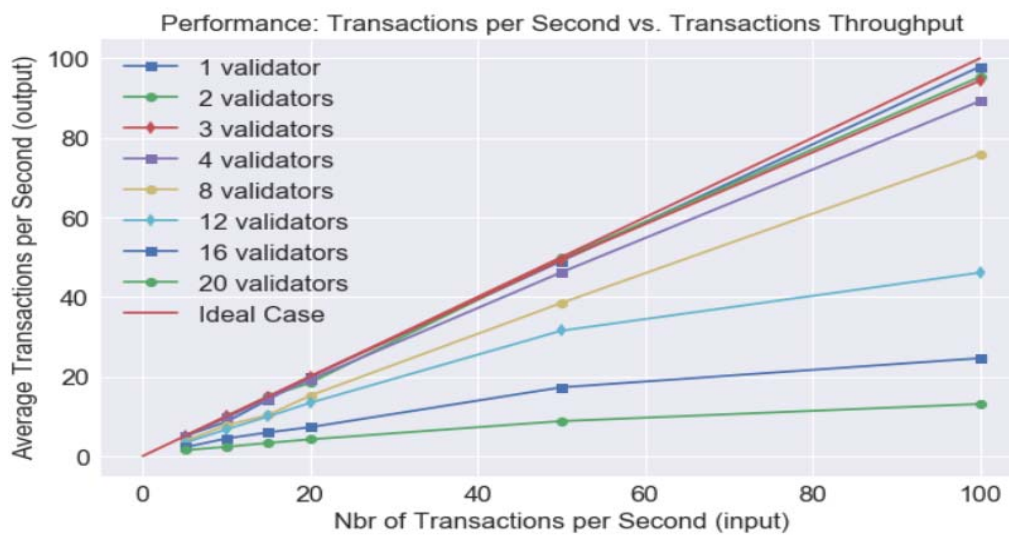

Figure 4: The number of transactions per second against validators.

From this figure, a consortium network with 1, 2, 3 or 4 validators can validate 100 transactions in a few seconds, which is close to the ideal case. However, for a consortium network comprised of 8, 12, 16 or 20 validators, the network requires several seconds to validate 100 transactions, which deviates far from the ideal situation.

**Performance Based On The Validation Time**
More time is required to valid a transaction as the number of validators increases in the consortium blockchain network, as shown in Figure 5. If a consortium network contains one validator, the average validation time is lower (less than 1 sec), however, validation time increases to almost 2 minutes when a network contains 20 validators in the experimental environment. Apparently, a compromise needs to be made between the desired network performance, the number of transactions sent per second, and the number of validating peers in a consortium blockchain network, which should be fully considered during the design stage. More specifically, an extra care must be taken to select the appropriate number of validators in the consortium network for a balance between security and transaction throughput.

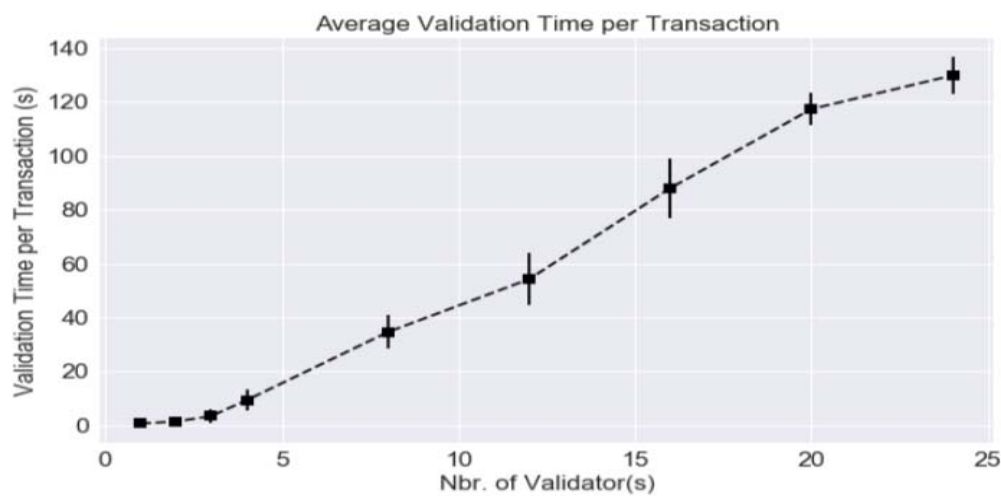

Figure 5: Validation time against the number of Validators.

**Evaluation On Security And Privacy**
This section presents a qualitative evaluation on how the proposed consortium blockchain-based system can prevent common threats to security and privacy in e-government systems.

*Prevention against distributed denial of service (DDoS) attacks* – Usually occur when attackers flood online services with massive traffic with fake requests in order to render the service unavailable. Generally, DDoS attacks tend to consume a significant amount of bandwidth and resources until the service is down. In the proposed consortium blockchain-based system, no centralised server that can be a direct target for DDoS attacks. The decentralised nature of the proposed consortium blockchain will allocate data and bandwidth to the less overloaded peers in the network to absorb DDoS attacks when it happens.

*Authentication and authorisation attacks* – This may occur when the adversarial users try to take control of the consortium blockchain network so that they can authorise themselves or introduce fake nodes for authorising users whilst taking control of the network. This is impossible in the proposed system since all peers of the consortium network are pre-selected by an authorised entities from an e-government agency of a particular government beforehand.

*Threats to anonymity* – Anonymity in blockchain-based applications may open the door for criminals to carry out unauthorised activities (Elisa et al., 2018b). In the proposed system, the blockchain information is only available to the consortium peers and authorised users. Therefore, any adversary trying to set up an anonymous connection will be detected instantly since the validators verify all users who attempt to access to the information stored in the consortium network.
The advantages of the proposed e-government consortium blockchain are summarised in Table 1.

Table 1: Advantages of the consortium blockchain-based e-government network.

| Advantage | Description(s) |
| --- | --- |
| Fast transaction speed | Only a selected group of participants process transactions. |
| High scalability | New participants can be added to a controlled number of consortium nodes when needed. |
| Low transaction costs | The transactions validators process transaction without incentives like in public blockchain. |
| Low energy consumption | A voting mechanism is used to pre-select the validating nodes, no computation is needed. |
| Low risk of 51% security attack | Random participants are not allowed to join the consortium unlike in public blockchain. |
| High transparency | Participants of the network know their peers within the consortium (more of enterprise). |
| High data integrity | Information is maintained in a consistent manner within the consortium network. |
| High collaboration | Different department and agencies will be able to share information on demand. |

## CONCLUSION

This work proposes a secure and privacy-preserving e-government architecture based on consortium blockchain technology for effective information sharing. The performance evaluation on a single transaction time, the number of transactions per second, and the number of validators has shown that, the consortium blockchain can be adopted in e-government systems by carefully keeping fewer number of validators in the network. Additionally, the proposed architecture can offer advantages such as high scalability, high transaction speed, high data integrity, high collaboration, low risk of 51% security attack, low energy consumption, low transaction cost and anonymity, whilst simultaneously ensuring the required level of security and privacy in e-government systems. The active future work is to use Cyber-Defense Technology Experimental Research Laboratory (DETERLab) test bed environment to evaluate the proposed consortium blockchain-based e-government system for real time attacks.

## ACKNOWLEDGMENT

This work has been supported by the Commonwealth Scholarship Commission (CSCTZCS-2017-717) and the Royal Academy of Engineering (IAPP 1-100077).

## REFERENCES

Biswas, K., & Muthukkumarasamy, V. (2016). Securing smart cities using blockchain technology. In *2016 IEEE 18th international conference on high performance computing and communications; IEEE 14th international conference on smart city; IEEE 2nd international conference on data science and systems (HPCC/SmartCity/DSS)* (pp. 1392-1393). IEEE.
Buterin, V. (2014). A next-generation smart contract and decentralized application platform. *white paper, 3*(37).
Cachin, C. (2016). Architecture of the hyperledger blockchain fabric. In *Workshop on distributed cryptocurrencies and consensus ledgers , 310*, 4.
Zheping Huang &  Zen Soo(2019). Data leak exposes 364 million Chinese social media profiles tracked by police surveillance programme,security researcher says, South China Morning Post, Retrieved from http://scmp.com/tech/enterprises /article/2188662/ data-leak-exposes-364-million-chinese-social-media-profiles-tracked/.( accessed  27 September 2019).
Dib, O., Brousmiche, K.-L., Durand, A., Thea, E., & Hamida, E. B. (2018). Consortium blockchains: Overview, applications and challenges. *International Journal On Advances in Telecommunications, 11*(1&2).
Dorri, A., Kanhere, S. S., Jurdak, R., & Gauravaram, P. (2017). Blockchain for IoT security and privacy: The case study of a smart home. In *2017 IEEE international conference on pervasive computing and communications workshops (PerCom workshops)* (pp. 618-623). IEEE.
Elisa, N. (2017). Usability, accessibility and web security assessment of e-government websites in tanzania. *International Journal of Computer Applications, 164*,42-48.
Elisa, N., Li, J., Zuo, Z., & Yang, L. (2018a). Dendritic cell algorithm with fuzzy inference system for input signal generation. In *UK Workshop on Computational Intelligenc ,Springer ,*pp. 203–214.
Elisa, N., Yang, L., Chao, F., & Cao, Y. (2018b). A framework of blockchainbased secure and privacy-preserving e-government system. *Wireless Networks,* 1–11.
Elisa, N., Yang, L., Fu, X., & Naik, N. (2019). Dendritic cell algorithm enhancement using fuzzy inference system for network intrusion detection. In *2019 IEEE International Conference on Fuzzy Systems (FUZZ-IEEE)*, pp 1–6.

Elisa, N., Yang, L., & Naik, N. (2018, July). Dendritic cell algorithm with optimised parameters using genetic algorithm. In *2018 IEEE Congress on Evolutionary Computation (CEC)*. IEEE. pp. 1-8.

IPP Media (2018). How Tanzania lost Tanzanian shillings 187 billions to cyber criminals in 2016. Retrieved from https://www.ippmedia.com/en/business/ how-tanzania-lost-187bn-cyber-criminals-2016/ ( accessed  28 May 2018).

Karokola, G. R. (2012). A framework for securing e-government services: the case of Tanzania. PhD thesis, Department of Computer and Systems Sciences, Stockholm University.

Lambrinoudakis, C., Gritzalis, S., Dridi, F., and Pernul, G. (2003). Security requirements for e-government services: a methodological approach for developing a common pki-based security policy. Computer Communications, 26(16),1873-1883.

Mingxiao, D., Xiaofeng, M., Zhe, Z., Xiangwei, W., & Qijun, C. (2017). A review on consensus algorithm of blockchain. In *2017 IEEE International Conference on Systems, Man, and Cybernetics (SMC)*. IEEE, pp. 2567-2572

Nakamoto, S. (2008). Bitcoin: A Peer-to-Peer Electronic Cash System,pp.1-9.

Peterson, K., Deeduvanu, R., Kanjamala, P., & Boles, K. (2016). A blockchainbased approach to health information exchange networks. In *Proc. NIST Workshop Blockchain Healthcare*, *1*,1-10.

Ramya, U. M., Sindhuja, P., Atsaya, R. A., Dharani, B. B., & Golla, S. M. V. (2018). Reducing forgery in land registry system using blockchain technology. In *International Conference on Advanced Informatics for Computing Research* ,Springer, pp. 725-734.

Stefanova, K., Kabakchieva, D., & Borthwick, L. (2006). Innovative approach to identity management solution development for e-government at EU level. *Journal of Telecommunications and Information Technology,* 24-31.

Tian, F. (2016, June). An agri-food supply chain traceability system for China based on RFID & blockchain technology. In *2016 13th international conference on service systems and service management (ICSSSM)*, pp. 1-6..

Turkanović, M., Hölbl, M., Košič, K., Heričko, M., & Kamišalić, A. (2018). EduCTX: A blockchain-based higher education credit platform. *IEEE access*, *6*, 5112-5127..

UK Cyber-Attacks (2019). Cyber Security Breaches Survey 2019. https://www.gov.uk/government/statistics/cyber-security-breaches-survey-2019/. ( accessed 27 September 2019).

UN, E. (2018). United nations e-government survey 2018: Gearing e-government to support transformation towards sustainable and resilient societies. New York, NY: United Nations.

US Cyber-Attacks (2019). List of data breaches and cyber attacks in April 2019 – 1.34 billion records leaked. https://www.itgovernance.co.uk/blog/list-of-data-breaches-and-cyber-attacks-in-april-2019-1-34-billion-records-leaked/. (accessed 22 September 2019).

Wireless Broadband (2018). Wireless Broadband Alliance Launches City Wi-Fi Roaming Project. Retrieved from https://wballiance.com/wireless-broadband-alliance-launches-city-wi-fi-roaming-project/

Wolrich, G. M., Yap, K. S., Guilford, J. D., Gopal, V., & Gulley, S. M. (2014). Instruction set for message scheduling of sha256 algorithm. *US Patent* 8,838,997.

Yang, L., Elisa, N., & Eliot, N. (2019). Privacy and security aspects of E-government in smart cities. In *Smart cities cybersecurity and privacy*. Elsevier, pp. 89-102

Yang, L., Li, J., Fehringer, G., Barraclough, P., Sexton, G., & Cao, Y. (2017, July). Intrusion detection system by fuzzy interpolation. In *2017 IEEE international conference on fuzzy systems (FUZZ-IEEE)* pp. 1-6